\documentclass[final,12pt, times]{elsarticle}

\usepackage{amssymb}
\usepackage{amsmath}
\journal{arXiv}
\usepackage{lineno}
\usepackage{color}

\begin{document}

\begin{frontmatter}

\title{A Novel Multi-Objective Reinforcement Learning Algorithm for Pursuit-Evasion Game}
\author[a]{Penglin Hu}
\author[a]{Chunhui Zhao*}
\ead{zhaochunhui@nwpu.edu.cn}
\author[a]{Quan Pan}%% Author name

\affiliation[a]{organization={School of Automation},%Department and Organization
            addressline={Northwestern Polytechnical University}, 
            city={Xi'an},
            postcode={710129}, 
            state={Shaanxi},
            country={China}}

%% Abstract
\begin{abstract}
In practical application, the pursuit-evasion game (PEG) often involves multiple complex and conflicting objectives. The single-objective reinforcement learning (RL) usually focuses on a single optimization objective, and it is difficult to find the optimal balance among multiple objectives. This paper proposes a three-objective RL algorithm based on fuzzy Q-learning (FQL) to solve the PEG with different optimization objectives. First, the multi-objective FQL algorithm is introduced, which uses the reward function to represent three optimization objectives: evading pursuit, reaching target, and avoiding obstacle. Second, a multi-objective evaluation method and action selection strategy based on three-dimensional hypervolume are designed, which solved the dilemma of exploration-exploitation. By sampling the Pareto front, the update rule of the global strategy is obtained. The proposed algorithm reduces computational load while ensuring exploration ability. Finally, the performance of the algorithm is verified by simulation results.
\end{abstract}


% \begin{highlights}
% \end{highlights}

%% Keywords

\end{frontmatter}


% \linenumbers


\section{Introduction}\label{sec:introduction}
The pursuit-evasion game (PEG) has been extensively studied and applied in both military and civilian fields \cite{Weintraub2020}. With the development of computer technology, reinforcement learning (RL) has demonstrated remarkable effectiveness in solving PEG problems. RL enables an agent to continuously interact with the environment. By using rewards and observation results, it gradually updates and optimizes the agent's training process, thereby finding the optimal strategy for the PEG \cite{Sutton}. 

For the multiple-to-single PEG problem, a distributed cooperative pursuit strategy based on RL has been developed. Using a centralized critic and distributed actors, as well as a learning-based communication mechanism, the complexity has been reduced and resources have been saved \cite{Wang2020-1}. The author utilizes RL and matrix game theory to solve the multiplayer PEG problem. The game process is expressed as a series of discrete matrix games, and min-max Q-learning is applied to generate the entries of the payoff matrix, thus obtaining the optimal actions of the players at each stage \cite{Selvakumar-2022}. In \cite{2024ji}, a cooperative pursuit strategy based on min-max Q-learning is proposed. By training both the pursuer and the evader simultaneously, the optimal adversarial strategy was obtained. Inspired by biology, the PEG has been explored from the perspective of dynamics. The RL algorithm is adopted to improve the pursuit-evasion strategy, enabling the faster pursuer to capture the evader successfully \cite{Xiong-2022}. The proposed model-free RL method obtains the optimal PEG strategy through online policy iteration and can achieve Nash equilibrium games without knowing the players' dynamics \cite{2024Jia}.

With the increasing maturity of research on PEG algorithms, researchers have begun to tackle PEG problems in practical applications. For the PEG of unmanned surface vehicles (USVs), an online RL algorithm has been proposed. The motion equations of the PEG are formulated as a differential game, which overcomes the weaknesses of data-driven learning \cite{Yongkang-2023}. For the PEG of USVs in restricted environments, a distributed algorithm based on RL is proposed, and the pursuit performance is enhanced through a reward function based on the artificial potential field method \cite{2023wang}. Aiming at the pursuit-evasion confrontation game problems among USVs in complex multi-obstacle environments, a confrontation game strategy combining RL and imitation learning algorithms is proposed \cite{2023Xiuqing}. Based on the decision making sequence of the pursuer and evader, a PEG of USVs grounded in Stackelberg game is proposed, and a globally balanced pursuit-evasion strategy is obtained by adopting the RL algorithm \cite{Xiaoxiang2024}.

In addition to the research on PEGs based on USVs, studies on PEGs based on unmanned aerial vehicles (UAVs) have also witnessed remarkable development. Combining feedforward control technology and the RL algorithm, a control scheme for the PEG of the UAV system is proposed, and the Nash equilibrium strategy of the PEG is obtained \cite{hang145}. A PEG algorithm combining differential game technology and integral reinforcement learning (IRL) is proposed. This algorithm is used to capture illegal UAVs and safeguard the safety of industrial scenarios \cite{Qingxue2024}. In addition, for the intrusion and monitoring scenarios in public safety applications, an UAV pursuit-evasion control framework with swarm intelligence is proposed. Precise strategic planning is achieved by means of the RL algorithm \cite{Weiqiang2024}. Some researchers have applied the PEG to the satellite pursuit-evasion scenario. They obtained the motion strategy of the spacecraft based on the deep deterministic policy gradient (DDPG) algorithm, thus solving the PEG problem in elliptical orbits \cite{2024Weizhuo}.

Traditional RL algorithms provide effective solutions to the decision making problems of the PEG. However, they rely on precise state information and clear action selection to guide agents to learn optimal strategies. Nevertheless, many real world scenarios are filled with uncertainty, fuzziness, and imprecision. Fuzzy RL, by introducing the fuzzy set theory, enables agents to make relatively reasonable decisions even when the information is imprecise, and endows them with stronger adaptability and decision making capabilities. For the PEG scenario of territorial defense, a algorithm based on fuzzy actor-critic learning (FACL) is proposed. Precise control is achieved through multiple reward functions of different types \cite{Asgharnia693}. In the multiple-to-multiple territorial defense PEG scenario, a hierarchical RL algorithm is proposed, and active target defense is achieved through task assignment \cite{2022Asgharnia}. Combined with the FACL and Kalman filtering methods, the single-to-single PEG in continuous environments is solved \cite{Penglin-CAC}. In \cite{509}, a PEG strategy based on fuzzy Q-learning (FQL) is proposed. An artificial potential field based reward function is designed, which improves the learning performance of the algorithm.

In summary, RL algorithms have achieved remarkable results in PEG problems. However, the learning algorithms in the aforementioned literature all adopt a single-objective optimization approach, which is to maximize the cumulative reward. However, in practical applications, PEG problems often involve multiple conflicting objectives. By assigning weights to and optimizing multiple objectives, a more balanced and efficient pursuit-evasion strategy can be found, improving the robustness and adaptability of the algorithm. As far as we know, there is no multi-objective RL algorithm yet to address PEG problems. Therefore, this paper proposes a three-objective RL algorithm based on FQL. Three optimization objectives, namely evading pursuit, reaching target, and avoiding obstacle, are considered. The dilemma of exploration-exploitation is solved by an action selection strategy based on three-dimensional hypervolume. After obtaining the Pareto front, the solution on the Pareto front is selected through the uniform sampling method for global strategy update, which ensures the learning performance of the algorithm and greatly reduces the computational load, thus avoiding the curse of dimensionality. The structure of the paper is as follows: 

Section \ref{2} introduces the model of PEG and the FQL algorithm. Section \ref{3} presents the multi-objective FQL algorithm, the three-dimensional hypervolume evaluation method, and the global strategy update. Section \ref{4} verifies the ability to obtain the Pareto front based on hypervolume through simulation, as well as the effects of the PEG under different learning parameters. Section \ref{5} summarizes the paper.

%%%%%%%%%%%%%%%%%%%%%%%%%%%%%%%%%%%%%%%%%%%%%%%%%%%%%%%%%%%%%%%%%%%%%%%%%%%%%%%%%%%%%%%%%%%%%%%%%%%%%%%%%%%%%%%%%%%%%%%%%%%%%%%%%%%%
\section{Pursuit-evasion game and fuzzy Q-learning}\label{2}
\subsection{The model of pursuit-evasion game}
In the PEG, the objective of the evader is to reach the target area while avoiding being captured by the pursuer and colliding with obstacles. Both sides of the game only know the position information of their opponents, but not the strategies they adopt. Define the model of the agent as
\begin{equation}
\label{model-MOFQL}
\begin{aligned}
\dot x &= v \cdot \cos \beta \\
\dot y &= v \cdot \sin \beta\\
\dot \beta &= \frac{{v \cdot \psi }}{L},
\end{aligned}
\end{equation}
where $(x, y)$ represents the position of the agent, $v$ is the velocity, $\beta$ is the heading angle, $L$ is the wheelbase, and $\psi$ is the steering angle. To satisfy the constraints of the model, the steering angle satisfies $\psi  \in \left[ { - \frac{\pi }{3},\frac{\pi }{3}} \right]$. The input of the PEG system consists of the following parameters
\begin{equation}
\label{input_state}
\mathbf{Inputs} = \left[ {{d_{ET}},{d_{EP}},{d_{EO}},{\beta _{E}}} \right],
\end{equation}
where $d_{ET}$ represents the distance between the evader and the target, $d_{EP}$ represents the distance between the evader and the pursuer, and $d_{EO}$ represents the distance between the evader and the obstacle. ${\beta _{E}}$ is the angular difference between the evader's heading and the $x$-axis. The PEG is judged to end when the evader reaches the target area or is captured by the pursuer.

In the PEG studied in this paper, taking the evader as the research object, there are three interrelated optimization objectives to consider. Objective 1: (evading pursuit) The evader needs to increase the distance from the pursuer; Objective 2: (reaching target) The evader needs to reduce the distance from the target point; Objective 3: (avoiding obstacle) The evader needs to increase the distance from the obstacle. These optimization objectives are described by the reward functions
\begin{equation}
\label{mo-reward}
\begin{aligned}
{r_{EP}} &= {r_{EP}}(t + 1) - {r_{EP}}(t)\\
{r_{ET}} &= {r_{ET}}(t) - {r_{ET}}(t + 1)\\
{r_{EO}} &= {r_{EO}}(t + 1) - {r_{EO}}(t).
\end{aligned}
\end{equation}
Define the overall reward function as
\begin{equation}
\label{R-PEG}
r = {k_1}{r_{EP}} + {k_2}{r_{ET}} + {k_3}{r_{EO}},
\end{equation}
where $k_1$, $k_2$ and $k_3$ represent the adjustment coefficients of the optimization objectives, respectively. By designing the overall reward function \eqref{R-PEG}, the evader can achieve multiple optimization objectives. Due to the conflicts existing among these objectives, a balance needs to be struck among them, which can be achieved by adjusting the coefficients $k_1$, $k_2$ and $k_3$. However, it is impractical to iterate through all coefficient combinations one by one. Therefore, multi-objective RL can be utilized to optimize the agent's strategy, enabling it to select the best actions to optimize multiple objectives and improve learning efficiency.

\subsection{Fuzzy Q-learning}

In the Q-learning algorithm, the challenges of continuous spaces and the storage and update efficiency of the Q-table must be considered. The FQL algorithm, which generates global continuous actions for agents based on a predefined discrete action set, can effectively solve the above problems. Suppose the agent has $n$ inputs $\bar x = [{x_1},...,{x_n}]$, $m$ actions $A = \{ {a_1},...,{a_m}\}$, and $L$ fuzzy rules. During the training process, a reasonable action selection mechanism needs to be designed to overcome the exploration-exploitation dilemma. In the Softmax exploration mechanism, the higher the Q-value of an action, the higher the probability of it being selected. For rule $l \in L$, the probability of selecting action $a^l$ is
\begin{equation}
\label{Boltzmann-action}
\Pr ({a^l}) = \frac{{\exp \left( {\tau  \cdot Q(l,{a^l})} \right)}}{{\sum\limits_{k = 1}^{|A|} {\exp \left( {\tau  \cdot Q(l,{a^k})} \right)} }},
\end{equation}
where $Q(l,a^l)$ is the Q-value of action $a^l$ for a given rule $l$, $|A|$ is the size of the action space, and $\tau$ is the Softmax temperature. After selecting an action for each rule, the global action at time $t$ is
\begin{equation}
\label{global continuous action}
{a_t}({\bar x_t}) = \frac{{\sum\limits_{l = 1}^L {\left( {(\prod\limits_{i = 1}^n {{\mu ^{F_i^l}}({x_i})} ) \cdot {a^l}} \right)} }}{{\sum\limits_{l = 1}^L {\left( {\prod\limits_{i = 1}^n {{\mu ^{F_i^l}}({x_i})} } \right)} }}=\sum\limits_{l = 1}^L {\Phi _t^l a_t^l},
\end{equation}
where ${\Phi _t^l}$ is the activation intensity of rule $l$ at time $t$
\begin{equation}
\label{firing strength for l}
{\Phi _t^l} = \frac{{\prod\limits_{i = 1}^n {\mu ^{F_i^l}\left( {{x_i}} \right)} }}{{\sum\limits_{l = 1}^L {\left( {\prod\limits_{i = 1}^n {\mu ^{F_i^l}\left( {{x_i}} \right)} } \right)} }},
\end{equation}
where $\mu ^{F_i^l}$ is the membership degree of the fuzzy set ${F_i^l}$, which can be calculated by the Gaussian membership function or the triangular membership function. The global Q-function is 
\begin{equation}
\label{global Q-function}
{Q_t}({\bar x_t}) = \sum\limits_{l = 1}^L {\Phi _t^l{Q_t}(l,a_t^l)},
\end{equation}
where ${Q_t}(l,a_t^l)$ is the Q-value after performing the operation $a_t^l$ on rule $l$ at time step $t$. The global Q-function with the maximum Q-value is 
\begin{equation}
\label{maximum Q-value}
Q_t^*({\bar x_t}) = \sum\limits_{l = 1}^L {\Phi _t^l\mathop {\max }\limits_{a \in A} {Q_t}(l,a)}.
\end{equation}
The temporal difference (TD) error is
\begin{equation}
\label{TD errors for l}
{\tilde \varepsilon _{t + 1}} = {r_{t + 1}} + \gamma Q_t^*({\bar x_{t + 1}}) - {Q_t}({\bar x_t}),
\end{equation}
where $\gamma$ is the discount factor, $r_{t + 1}$ is the reward at time $t$, and the update rule of the Q-function is
\begin{equation}
\label{update for the Q}
\begin{aligned}
{Q_{t + 1}}(l,a_t^l) &= {Q_t}(l,a_t^l) + \alpha {{\tilde \varepsilon }_{t + 1}}\Phi _t^l\\
&= {Q_t}(l,a_t^l) + \alpha \left[ {{r_{t + 1}} + \gamma Q_t^*(l,a_{t + 1}^l) - {Q_t}(l,a_t^l)} \right]\Phi _t^l,
\end{aligned}
\end{equation}
where $\alpha$ is the learning rate.

\section{Multi-objective fuzzy Q-learning}\label{3}
\subsection{Single-objective RL and multi-objective RL}
The difference between single-objective RL and multi-objective RL lies in the change of optimization objectives, and this difference is mainly reflected in the difference of the reward function. In a single-objective RL, the reward is a scalar, expressed as $r_t = r$. However, in a multi-objective RL, the reward is a vector with $n$ objective signals, expressed as ${\vec r}_t = [r_1,r_2,...,r_n]^\top$. Therefore, in a single-objective RL, the Q-value of each action under a given state is a scalar. In contrast, in a multi-objective RL, the Q-value of each action under a given state is a set of vectors. Two numerical examples are used to analyze in detail the differences between single-objective RL and multi-objective RL in discrete states.
\begin{figure}[!ht]
	\centering
	\includegraphics[width=1\columnwidth]{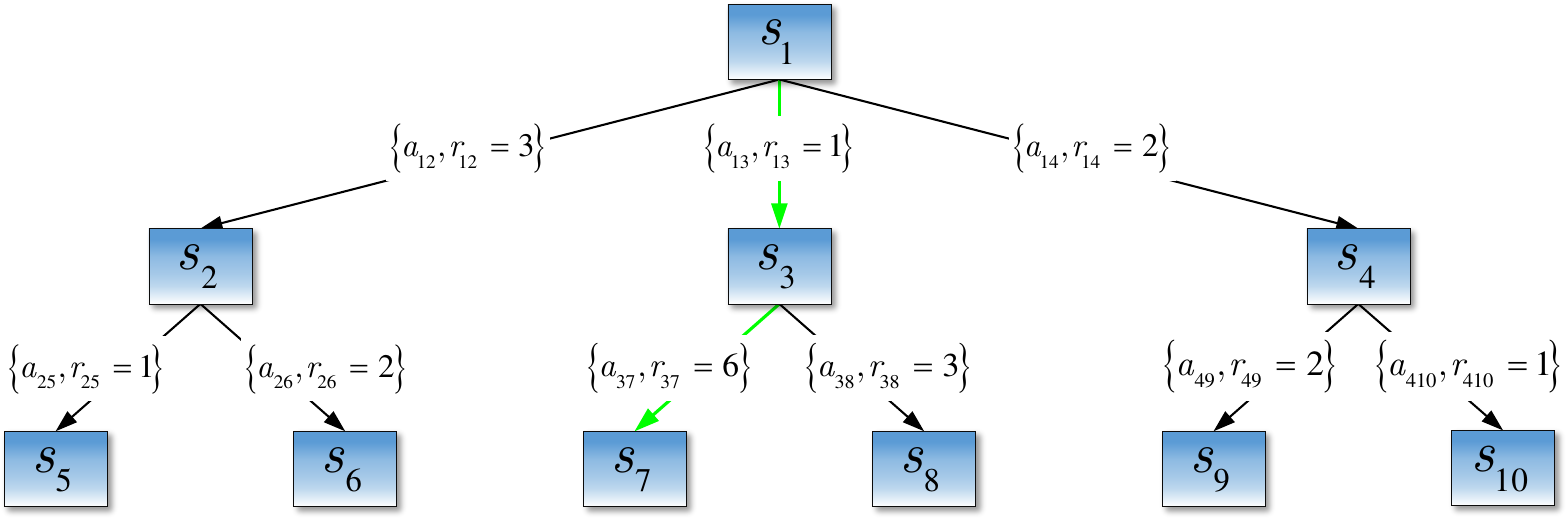}
	\caption{The single-objective reinforcement learning process.}
    \label{single_RL}
\end{figure}
\begin{figure}[!ht]
	\centering
	\includegraphics[width=1\columnwidth]{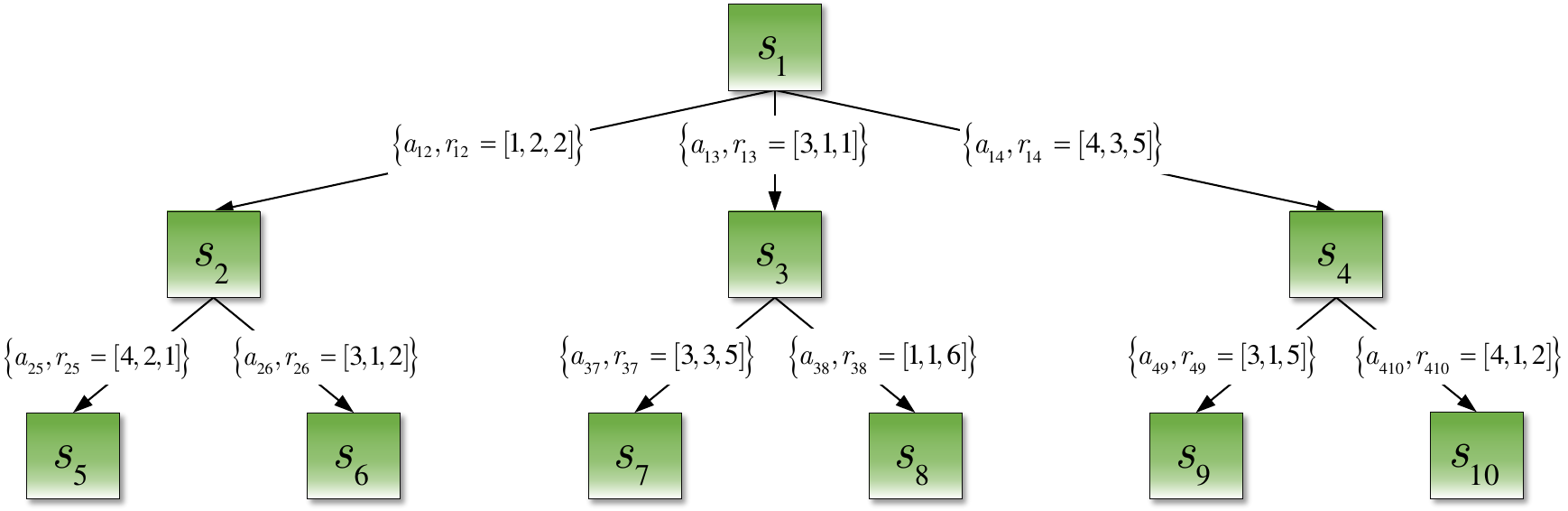}
	\caption{The multi-objective reinforcement learning process.}
	\label{multi_RL}
\end{figure}

\textbf{Example 1:} Fig. \ref{single_RL} demonstrates the steps of a single-objective RL, where the agent executes action $a_{ij}$, transitions from state $s_i$ to $s_j$ and obtains a reward ${r_{ij}}$. Let $V\left({{s_5}}\right) = V\left({{s_6}} \right) = \cdots = V\left({{s_{10}}} \right) = 0$. According to the Bellman equation ${V^*}({s_t}) = \mathop {\max }\limits_{{a_t} \in {A}} \left( {{r_{t + 1}} + \gamma {V^*}({s_{t + 1}})} \right)$, with $\gamma = 0.8$, calculate the optimal value function
\begin{equation}
\label{single-Bellman}
\begin{aligned}
{V^*}({s_2}) &= \max \left( {{r_{25}} + \gamma {V^*}({s_5}),{r_{26}} + \gamma {V^*}({s_6})} \right) = 2\\
{V^*}({s_3}) &= \max \left( {{r_{37}} + \gamma {V^*}({s_7}),{r_{38}} + \gamma {V^*}({s_8})} \right) = 6\\
{V^*}({s_4}) &= \max \left( {{r_{49}} + \gamma {V^*}({s_9}),{r_{410}} + \gamma {V^*}({s_{10}})} \right) =2\\
{V^*}({s_1}) &= \max \left( {{r_{12}} + \gamma {V^*}({s_2}),{r_{13}} + \gamma {V^*}({s_3}),{r_{14}} + \gamma {V^*}({s_4})} \right)\\
&= \max \left( {4.6,5.8,3.6} \right) = 5.8.
\end{aligned}
\end{equation}
Based on \eqref{single-Bellman}, the optimal strategy shown by the green line in Fig. \ref{single_RL} can be obtained. 
 
\textbf{Example 2:} Fig. \ref{multi_RL} demonstrates the steps of a multi-objective RL. In this example, the reward value becomes a $1\times3$ vector. Let $V\left({{s_5}}\right) = V\left({{s_6}} \right) =\cdots = V\left({{s_{10}}} \right) = [0,0,0]$, and calculate the optimal value function according to the Bellman equation
\begin{equation}
\label{multi-bellman}
\begin{aligned}
{V^*}({s_2}) &= \max \left( {{r_{25}} + \gamma {V^*}({s_5}),{r_{26}} + \gamma {V^*}({s_6})} \right) \\
&= \max \left( {[4,2,1],[3,1,2]} \right)\\
{V^*}({s_3}) &= \max \left( {{r_{37}} + \gamma {V^*}({s_7}),{r_{38}} + \gamma {V^*}({s_8})} \right)\\
&= \max \left( {[3,3,5],[1,1,6]} \right)\\
{V^*}({s_4}) &= \max \left( {{r_{49}} + \gamma {V^*}({s_9}),{r_{410}} + \gamma {V^*}({s_{10}})} \right)\\
&= \max \left( {[3,1,5],[4,1,2]} \right).
\end{aligned}
\end{equation}
Unlike in \eqref{single-Bellman}, the operator $\max(\cdot)$ is of no significance in \eqref{multi-bellman}. Moreover, there is no comparability  between the vector $[4,2,1]$ and the vector $[3,1,2]$. In practical applications, a non-dominant solution needs to be found based on the preference for a certain objective instead of using a greedy strategy to find the maximum value. Therefore, the Bellman equation is extended to a generalized form
\begin{equation}
\label{generalized-bellman}
{V^*}(s) = {\rm{ND}}\left( {\bigcup\limits_{s' \in {S}} {\left( {\vec r + \gamma {V^*}(s')} \right)}} \right),
\end{equation}
where $s'$ represents the next state, ${S}$ represents the agent's state space, and $\vec{r}$ represents the reward function in vector form. $ND(\cdot)$ is a function that returns the set of non-dominated vectors. Specifically, the $ND(\cdot)$ operator eliminates the dominant solutions considering the developer's preferences and obtains solutions with greater returns. Suppose the first objective is prioritized to determine the action-selection strategy and the dominant solutions are eliminated, then the value function in Example 2 can be defined as follows
\begin{equation*}
\label{multi-bellman1}
\begin{aligned}
{V^*}({s_2}) &= {\rm{ND}}\bigcup \left( {{r_{25}} + \gamma {V^*}({s_5}),{r_{26}} + \gamma {V^*}({s_6})} \right)={[4,2,1]}\\
{V^*}({s_3}) &= {\rm{ND}}\bigcup \left( {{r_{37}} + \gamma {V^*}({s_7}),{r_{38}} + \gamma {V^*}({s_8})} \right)= {[3,3,5]} \\
{V^*}({s_4}) &= {\rm{ND}}\bigcup \left( {{r_{49}} + \gamma {V^*}({s_9}),{r_{410}} + \gamma {V^*}({s_{10}})} \right)= {[4,1,2]}.
\end{aligned}
\end{equation*}
Therefore, the optimal value function ${V^*}({s_1})$ can be calculated
\begin{equation*}
\begin{aligned}
{V^*}({s_1}) &={\rm{ND}}\bigcup {\left( {{r_{12}} \!+\! \gamma {V^*}({s_2}),{r_{13}} \!+\! \gamma {V^*}({s_3}),{r_{14}} \!+\! \gamma {V^*}({s_4})} \right)} \\
&=\! {\rm{ND}} \bigcup {\left( {[4.8,3.6,2.6],[5.4,3.8,5.8],[7.2,3.4,6]} \right)} \\
&=\! {[[5.4,3.8,5.8],[7.2,3.4,6]]^ \top }.
\end{aligned}
\end{equation*}

Through the above examples, it can be seen that in a single-objective RL, the Q-value corresponding to each action of each rule is a scalar. In a multi-objective RL, the Q-value corresponding to each action of each rule is a set of non-dominated vectors, which is defined as follows
\begin{equation}
\label{QQ}
{\bf{q}}(l,a) = {\left[ {\begin{array}{*{20}{c}}
		{q_1^1}&{q_1^2}& \cdots &{q_1^n}\\
		{q_2^1}&{q_2^2}& \cdots &{q_2^n}\\
		\vdots & \vdots & \ddots & \vdots \\
		{q_k^1}&{q_k^2}& \cdots &{q_k^n}
		\end{array}} \right]_{k \times n}},
\end{equation}
where $n$ is the number of objectives to be optimized, and $k$ represents the number of non-dominated Q-values assigned to action $a$ under rule $l$. In this paper, three optimization objectives are considered, and assuming that there are four non-dominated Q-values for rule $l$ and action $a$, ${\bf{q}}(l,a)$ can be expressed as
\begin{equation}
\label{QQ2}
{\bf{q}}(l,a) = {\left[ {\begin{array}{*{20}{c}}
		{q_1^1}&{q_1^2}&{q_1^3}\\
		{q_2^1}&{q_2^2}&{q_2^3}\\
		{q_3^1}&{q_3^2}&{q_3^3}\\
		{q_4^1}&{q_4^2}&{q_4^3}
		\end{array}} \right]_{4 \times 3}}.
\end{equation}
In \eqref{QQ2}, the matrix elements are scalars and can be obtained by the Bellman equation.

In the proposed multi-objective FQL algorithm, the extended Bellman equation is defined as follows
\begin{equation}
\label{k-bellman}
\hat Q(s,a) = \vec r(s,a) \oplus \gamma N{D_t}(s,a),
\end{equation}
where $\hat Q(s,a)$ represents the Q-value of state $s$ and action $a$, $\vec r$ is the vector reward function. The symbol $\oplus$ represents an operation between a vector and a set of vectors, which is defined as follows
\begin{equation}
\label{suanzi}
v \oplus U = \bigcup\limits_{v'  \in U} {(v + v' )}.
\end{equation}
A numerical example is given as follows
\begin{equation*}
v \oplus U = \left[ {\begin{array}{*{20}{c}}
	1&2&4
	\end{array}} \right] \oplus \left[ {\begin{array}{*{20}{c}}
	2&3&1\\
	3&5&2\\
        2&2&1
	\end{array}} \right] = \left[ {\begin{array}{*{20}{c}}
	3&5&5\\
	4&7&6\\
        3&4&5
	\end{array}} \right].
\end{equation*}
In a single-objective FQL, as shown in \eqref{global Q-function}, the global Q-function maps the state to a scalar by selecting the maximum Q-value of each action given each rule to represent the quality of the state. In a multi-objective RL, for each action, there is more than one non-dominated Q-value. Therefore, in the multi-objective FQL, there will be more than one global Q-function. Hence, the Bellman equation needs to be used multiple times so that all global Q-functions are taken into account.

\begin{figure}[!ht]
	\centering
	\includegraphics[width=0.6\columnwidth]{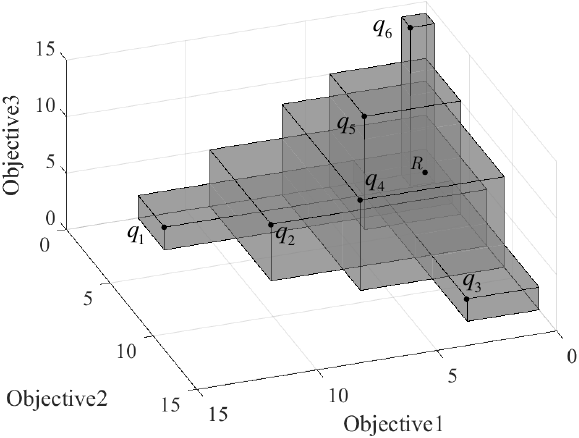}
	\caption{Hypervolume of the three-objective optimization problem.}
	\label{HV-shiyitu2}
\end{figure}

\subsection{Multi-objective evaluation based on three-dimensional hypervolume}
This paper adopts a three-dimensional hypervolume method to evaluate strategies, aiming to resolve the exploration-exploitation dilemma. The hypervolume refers to the volume enclosed by the points on the Pareto front and the selected reference point. In a two-objective optimization problem, it is the area, and in the three-objective optimization problem of this paper, it is the volume. Fig. \ref{HV-shiyitu2} shows the hypervolume of a three-objective problem with the reference point $R(0, 0, 0)$. The hypervolume represented in the figure is the Q-value $q(l, a)$ of the action corresponding to the given rule $l$. Assume that the non-dominated matrix ${\bf q}(l,a)$ is 
\begin{equation}
\label{FLSA}
{\bf q}(l,a) = {\left[ {\begin{array}{*{20}{c}}
		{12}&3&2\\
		9&7&5\\
		3&{13}&2\\
		6&9&8\\
		4&4&{10}\\
		1&1&{14}
		\end{array}} \right]_{6 \times 3}}.
\end{equation}
The hypervolume of action $a_i$ is defined as
\begin{equation}
\label{HV-computer}
{\cal H}_{{a}}^l = \prod\limits_{i = 1}^3 {q_i^l}  = q_1^l \cdot q_2^l \cdot q_3^l,
\end{equation}
where ${q_i^l}$ is the Q-value corresponding to the optimization objective. By normalizing \eqref{HV-computer}, we can obtain
\begin{equation}
\label{HV-normalize}
\overline {{\cal H}}_{a}^l  = \frac{{{\cal H}_{{a}}^l}}{{\sum\limits_{i = 1}^{|A|} {{\cal H}_{{a_i}}^l} }},
\end{equation}
where $|A|$ represents the size of the action space. Based on the softmax function, the probability that action $a_j$ is selected as the output parameter by rule $l$ is
\begin{equation}
\label{Pr-softmax}
\Pr \left( {a_j^l} \right) = \frac{{\exp (\tau  \cdot \overline{{\cal H}} _{a_j}^l)}}{{\sum\limits_{i = 1}^{|A|} {\exp (\tau \cdot \overline{{\cal H}}_{a_i}^l)} }},
\end{equation}
where $\tau$ is the temperature of the softmax function. At each time step, the softmax function is used to calculate the probability of choosing the corresponding action, and then the action with the highest probability is selected to strike a balance between exploration and exploitation. Let $\tau = 1.0$, according to the data in Table \ref{tab1-1}, it can be seen that the probability of action $a_4$ being selected as the output parameter of the rule $l$ is the highest. The output parameters of the fuzzy logic controller are selected through the hypervolume method. Since the softmax function shown in \eqref{Pr-softmax} is used, the output parameters are a combination of exploitation and exploration. 

\begin{table}[!ht]
	\caption{Hypervolume, normalized value, probability of each action, where $\tau = 1.0$}
	\begin{center}
		\tabcolsep 4pt
		\begin{tabular}{cccc}
			\hline
			Action $ a_i $ & Hypervolume ${{\cal H}}_{a_i}^l  $ & {Normalized $ \overline {{\cal H}}_{a_i}^l $}  & {Probability $ \Pr \left( {a_i^l} \right)  $}\\
			\hline
			$a_1$ & 72 & 0.0672 & 0.1494\\
			$a_2$ & 315 & 0.2941  & 0.1875\\
			$a_3$ & 78 & 0.0728  & 0.1503\\
			$a_4$ & 432 & 0.4034  & 0.2091\\
			$a_5$ & 160 & 0.1494  & 0.1622\\
			$a_6$ & 14  & 0.0131  & 0.1415\\
			\hline
		\end{tabular}
		\label{tab1-1}
	\end{center}
\end{table}

\subsection{Global Q-function calculation}
Like in single-objective FQL, a global Q-function also needs to be calculated in multi-objective FQL. As shown in \eqref{maximum Q-value}, in single-objective FQL, the global Q-function is calculated based on the maximum Q-value of each action under a given rule. In multi-objective FQL, the actions under a given rule have a set of non-dominated Q-values. Given rule $l$, the global Q-function $Q_l^*$ is defined as the non-dominated union of all ${\bf q}(l, a)$ of action $a \in {A}$, as follows
\begin{equation}
\label{globle-Q}
Q_l^* = ND\left( {\bigcup\limits_{a \in {A}} {{\bf  q}(l,a)} } \right),
\end{equation}
where $Q_l^*$ is a $k\times n$ matrix, $k$ refers to the number of non-dominated Q-values, and $n$ is the number of objectives, which is three in this paper. 

After the operation of the $ND(\cdot)$ operator, the dominated solutions in the Q-values are respectively eliminated, so that the elements in $Q_l^*$ do not dominate each other, thus obtaining the union of the non-dominated solutions of the actions. However, after multiple updates, the dimension of $Q_l^*$ will be extremely large, resulting in a Pareto front composed of a large number of points. Theoretically, all members of the Pareto front should be used to calculate multiple global Q-functions. Nevertheless, this will bring a huge computational burden. Therefore, we uniformly select a finite number of elements from the members of $Q_l^*$ to approximate the Pareto front and obtain the global Q-function under the condition of reducing the computational load.

\begin{figure}[!ht]
	\centering
	\includegraphics[width=1\columnwidth]{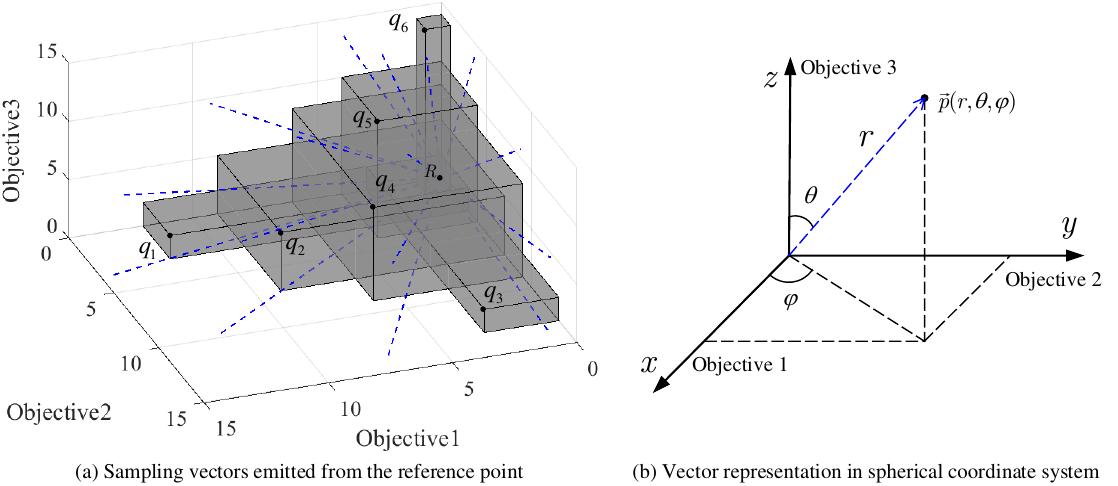}
	\caption{Schematic diagram of Pareto front sampling.}
	\label{HV-shiyitu-1}
\end{figure}

As shown in Fig. \ref{HV-shiyitu-1}(a), through the sampling method, starting from the reference point, $H$ rays are evenly drawn to sample the points on the Pareto front, obtaining a finite number of elements. As shown in Fig. \ref{HV-shiyitu-1}(b), in the spherical coordinate system, each ray is represented by a vector. The vector defined by point $p$ is $\vec p\left( {r,\theta,\varphi } \right)$, where $r$ is the radial distance. The angles $\theta$ and $\varphi$, which represent the angles between the vector and the coordinate axes, are uniformly distributed in the interval $[0,\frac{\pi}{2}]$. The angle between each ray and the coordinate axes reflects the influence of each objective. For example, the angle combination $\theta ={\frac{\pi}{2}}, \varphi = 0$ indicates that the strategy only pursues the completion of objective 1, the angle combination $\theta ={\frac{\pi}{2}}, \varphi = \frac{\pi}{2}$ indicates that the strategy only pursues the completion of objective 2, and the angle $\theta = 0$ indicates that the strategy only pursues the completion of objective 3. During the algorithm update process, $\tan \theta$ and $\tan \varphi$ are calculated to represent the degree of preference of the strategy for the objectives. 

\begin{figure}[!ht]
	\centering
	\includegraphics[width=0.6\columnwidth]{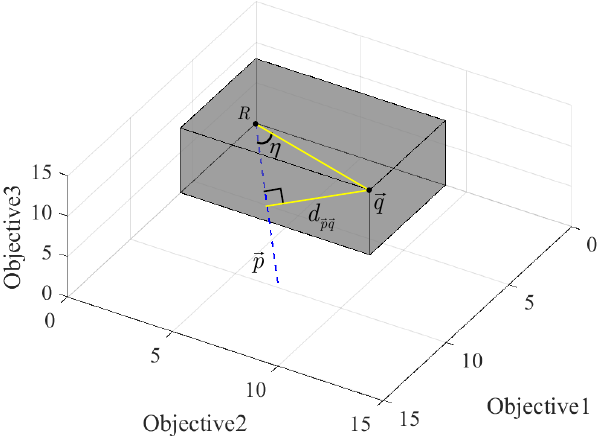}
	\caption{Select the point on the Pareto front.}
	\label{HV-shiyitu-2}
\end{figure}

During the algorithm update process, given rule $l$, the Q-value closest to the ray on the Pareto front is selected. As shown in Fig. \ref{HV-shiyitu-2}, the angle between the vector \(\vec p_i\left( {{r_i},{\theta _i},{\varphi _i}} \right)\) where the ray \( {p_i}\left( {i = 1,2,...,H} \right)\) is located and the vector \(\vec {q}_j\left( {{r_j},{\theta _j},{\varphi _j}} \right)\) where the point \( {q_j}\left( {j = 1,2,...} \right)\) on the Pareto front is located is 
\begin{equation}
\eta  = \arccos \left( {\cos {\theta _i}\cos {\theta _j} + \sin {\theta _i}\sin {\theta _j}\cos \left( {{\varphi _j} - {\varphi _i}} \right)} \right).
\end{equation}
The minimum distance from the point $q_j$ to the ray represented by the vector $\vec p_i$ is
\begin{equation}
{d_{\vec p_i \vec q_j}} = \left| {{r_j}\sin \eta } \right|.
\end{equation}
Repeat the above process for each rule \(l\in L\), and denote the selected Q-value as \( {G_{l({\theta _i},{\varphi _i})}^*} \). Each rule \(l\) has \(H\) points, and each point is associated with the slope \(({{\theta _i},{\varphi _i}})\), and the tuple \(({{\theta _i},{\varphi _i}})\) represents the different weights of each strategy for the optimization objectives. The global Q-function with input \(\bar x\) is as follows
\begin{equation}
\label{quanju-Q}
Q_{({\theta _i},{\varphi _i})}^*\left( {\bar x} \right) = \sum\limits_{l = 1}^L {{\Phi ^l}\left( {\bar x} \right)G_{l({\theta _i},{\varphi _i})}^*} ,\quad i = 1,2,...,H,
\end{equation}
where ${G_{l({\theta _i},{\varphi _i})}^*}$ and $Q_{({\theta _i},{\varphi _i})}^*$ are $1\times3$ vectors.

\subsection{Global rule update}
In the single-objective FQL algorithm, the temporal difference is shown in \eqref{TD errors for l}. At each time step, the triggering rules are updated through the scalar reward function and \(Q_t\) in \eqref{update for the Q}. Since there is only a single \(Q_t\), the rule is updated only once within one time step. In the multi-objective FQL algorithm, multiple global Q-functions are defined based on the preferences of each objective according to the needs of players or tasks, that is, there are multiple optimal Q-functions. The triggered rules need to be updated for each $Q_{(\theta_i,\varphi_i)}^*$ respectively. After the update, the dominated Q-values are eliminated, and only the non-dominated Q-values are stored.

Before introducing the global rule update, the operation $\ominus$ is introduced, and the operation $\oplus$ in \eqref{suanzi} is extended to the operation between matrices, which is defined as follows
\begin{equation}
\label{oplus2}
\begin{array}{l}
{U_{n \times g}} \oplus {V_{n \times h}} = \bigcup\limits_{i = 1}^g {\bigcup\limits_{j = 1}^h {\bigcup\limits_{k = 1}^n {\left( {{U_{(k,i)}} + {V_{(k,j)}}} \right)} } } \\
{U_{n \times g}} \ominus {V_{n \times h}} = \bigcup\limits_{i = 1}^g {\bigcup\limits_{j = 1}^h {\bigcup\limits_{k = 1}^n {\left( {{U_{(k,i)}} - {V_{(k,j)}}} \right)} } }. 
\end{array}
\end{equation}
The following numerical example is given to explain the operation $\oplus$
\begin{equation*}
\label{example}
U \oplus V \!=\! \left[ {\begin{array}{*{20}{c}}
	{12}&2&3\\
	9&7&5
	\end{array}} \right] \oplus \left[ {\begin{array}{*{20}{c}}
	3&{13}&2\\
	6&9&8\\
	4&4&{10}
	\end{array}} \right] \!\!=\! \left[ {\begin{array}{*{20}{c}}
	{15}&{15}&5\\
	{18}&{11}&{11}\\
	{16}&6&{13}\\
	{12}&{20}&7\\
	{15}&{16}&{13}\\
	{13}&{11}&{15}
	\end{array}} \right].
\end{equation*}
It can be seen that the operation $\oplus$ sums the values at the corresponding positions. Similarly, the operation $\ominus$ subtracts the corresponding values.

The temporal difference error is defined as follows
\begin{equation}
\label{wu-cha}
{\tilde \varepsilon} _{(\theta_i ,\varphi_i )}^l(t + 1) = {\vec r_{t + 1}} + \gamma Q_{(\theta_i ,\varphi_i )}^*({\bar x_{t + 1}}) \ominus {{\bf q}_t}(l,{a^l}),
\end{equation}
where ${\tilde \varepsilon}_{(\theta_i,\varphi_i)}^l$ represents the difference error of the $i$-th $Q_{(\theta_i,\varphi_i)}^*({\bar x_{t + 1}})$ corresponding to rule $l$. $\vec r$ and $Q_{(\theta_i,\varphi_i)}^*$ are $1\times3$ vectors, ${{\bf q}_t}(l,{a^l})$ is an $k\times3$ matrix. Through the $\ominus$ operation, the $1\times3$ vector is added to each row of the matrix ${{\bf q}_t}(l,{a^l})$. Where ${{\bf q}_t}(l,{a^l})$ has the same dimension as ${\tilde \varepsilon}_{(\theta_i,\varphi_i)}^l$, and the update rule of ${{\bf q}_t}(l,{a^l})$ is as follows
\begin{equation}
\label{update-q}
{{\bf q}_{t + 1}}\left( {l,a_t^l} \right) \!=\! ND\!\left( {\bigcup\limits_{i = 1}^H {\left( {{{\bf q}_t}\left( {l,a_t^l} \right) \oplus \alpha  \cdot \tilde \varepsilon _{\left( {{\theta _i},{\varphi _i}} \right)}^l\left( {t + 1} \right) \cdot \Phi _t^l} \right)} } \right).
\end{equation}
In \eqref{update-q}, $H$ represents the number of selected rays, that is, the number of $Q_{(\theta_i,\varphi_i)}^*$. $\alpha$ is the learning rate, and $\Phi_t^l$ is the activation intensity of rule $l$ at time $t$. After the temporal difference error $\tilde \varepsilon _{\left( {{\theta _i},{\varphi _i}} \right)}^l$ is multiplied by the learning rate $\alpha$ and the activation intensity $\Phi_t^l$, it is added to the Q-value through the operation $\oplus$. Repeat the above process until all $H$ rays are traversed, and $H$ global Q-functions are obtained. Then, take the union of all Q-values, and obtain the non-dominated Q-values through the operation $ND(\cdot)$. Usually, the dimension of ${{\bf q}_{t + 1}}\left( {l,a_t^l} \right)$ is larger than that of ${{\bf q}_{t}}\left( {l,a_t^l} \right)$ because at each update, the union of multiple temporal difference errors is assigned to ${{\bf q}_{t + 1}}\left( {l,a_t^l} \right)$, so the dimension of the Q-value will grow exponentially. Therefore, by using the uniform sampling method to retain $H$ representative Q-values, this problem can be well solved. 

The proposed multi-objective FQL algorithm can find the non-convex regions of the Pareto front. The Pareto front is not discrete but a continuous surface. The algorithm uses discrete sampling to approximately estimate the continuous Pareto front. In addition, the algorithm can find multiple strategies in each execution without knowing the preferences for each objective. Since the multi-objective FQL follows the update process of classical Q-learning, its convergence can be guaranteed.

%%%%%%%%%%%%%%%%%%%%%%%%%%%%%%%%%%%%%%%%%%%%%%%%%%%%%%%%%%%%%%%%%%%%%%%%%%%%%%%%%%%%%%%%%%%%%%%%%%%%%%%%%%%%%%%%%%%%%%%%%%%%%%%%%%%%%%%%%%%%%%%%%%%%%%%%%%%%%%%%%%%%%%%%%%%%%%%%
\section{Simulation and analysis}\label{4}
\subsection{Pareto front simulation based on three-dimensional hypervolume}
To evaluate the performance of the proposed algorithm in obtaining the Pareto front, a random operator is used to generate different types of data points. Specifically, three types of data point sets are generated, corresponding to three geometric shapes: convex surface, plane, and concave surface. Within the space covered by each surface, 500 three-dimensional points are randomly sampled to simulate the Q-value situation generated during the algorithm learning process. 

\begin{figure}[!ht]
	\centering
	\includegraphics[width=0.95\columnwidth]{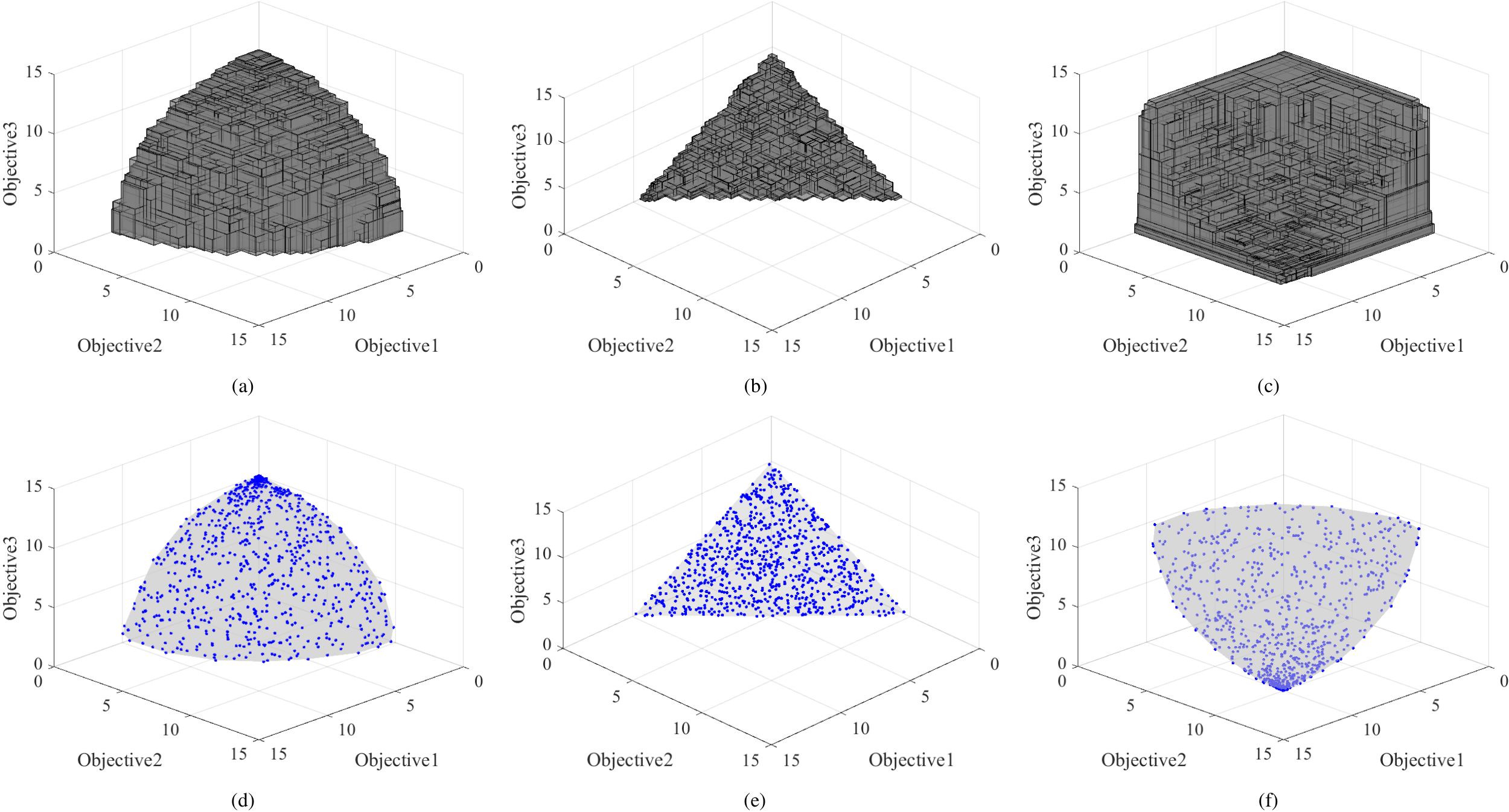}
	\caption{Hypervolume and Pareto front of different data types.}
	\label{different-3D1}
\end{figure}

As shown in Fig. \ref{different-3D1}, (a), (b) and (c) represent the three-dimensional hypervolumes represented by the sampled data points in the three-dimensional space, and (d), (e) and (f) represent the corresponding Pareto front results. It can be seen that the proposed algorithm obtains the hypervolumes composed of non-dominated solutions of each type and gets the Pareto front. The simulation results demonstrate  the rationality of \eqref{globle-Q}. Therefore, during the learning process, the designed algorithm can be used to obtain non-dominated solutions of the global Q-function.

\begin{figure}[!ht]
	\centering
	\includegraphics[width=0.9\columnwidth]{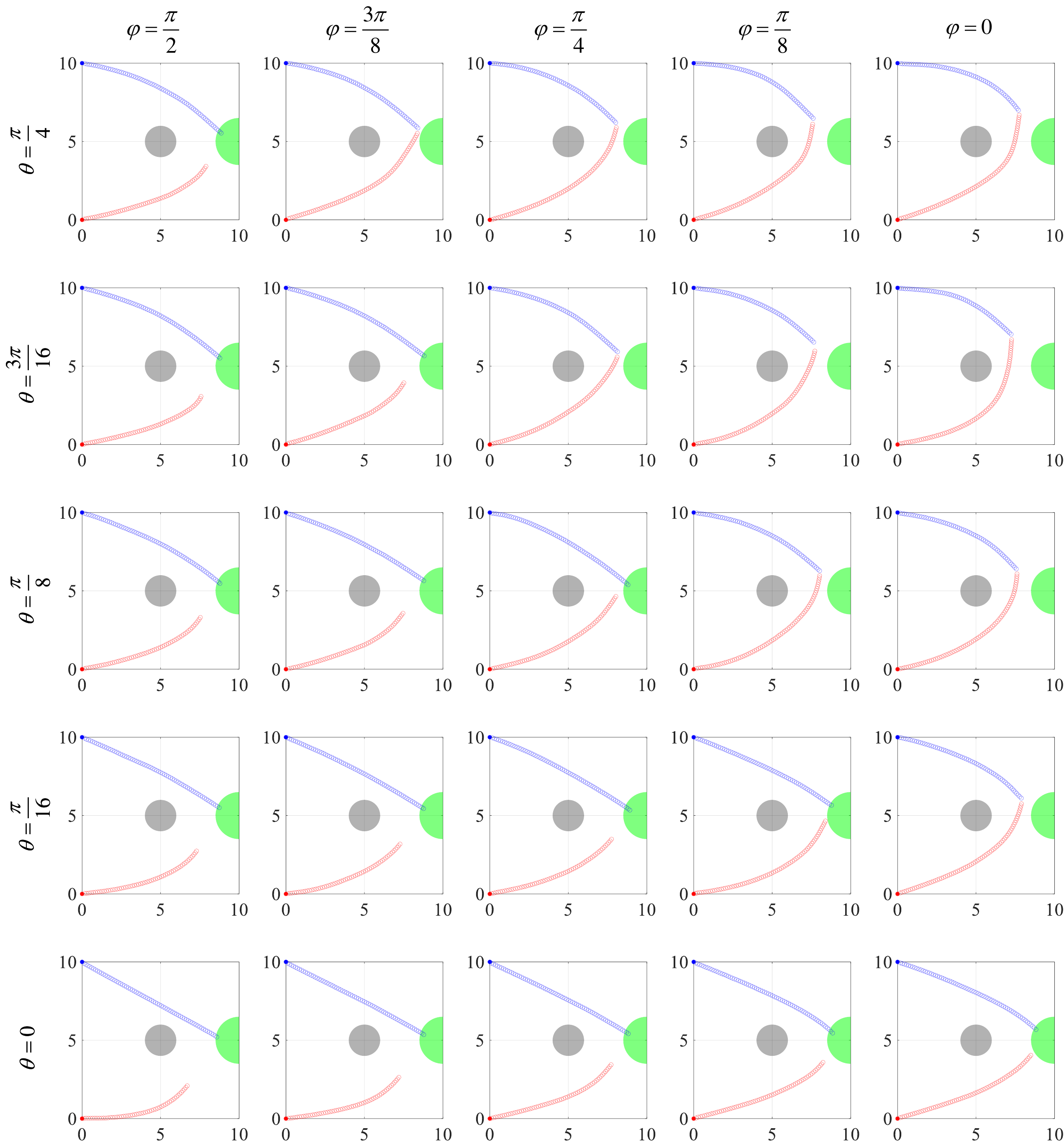}
	\caption{PEG trajectories under different sampling angles, the green area represents the target region, and the gray area represents the obstacle.}
	\label{morl-fql-guiji}
\end{figure}
%%%%%%%%%%%%%%%%%%%%%%%%%%%
\subsection{The simulation of PEG based on different learning parameters}
In the simulation experiment, the speed of the pursuer is set to $1.1\rm{m/s}$, the speed of the evader is $1.0\rm{m/s}$. The size of the PEG scenario is a square area of $10{\rm{m}} \times 10{\rm{m}}$. The target area is a semi-circular area with the center at $(10, 5)$ and a radius of $2\rm{m}$, and the obstacle is located at $(5, 5)$. To reduce the computational load, the triangular membership function is selected. The input interval of distance is $[0,10]$, and each input has 6 membership degrees. The input interval of angle is $\left[ { - \frac{\pi }{4},\frac{\pi }{4}} \right]$, and each input has 5 membership degrees. During the simulation, the maximum number of episodes is set to $1000$, and the learning rate is $0.01$.

As depicted in Fig. \ref{HV-shiyitu-1}, the angle $\varphi$ represents the degree of preference for objective 1 (evading pursuit) and objective 2 (reaching target), while the angle $\theta$ represents the degree of preference for objective 3 (avoiding obstacle). Here, $\varphi$ takes values of $\{ \frac{\pi }{2},\frac{{3\pi }}{8},\frac{\pi }{4},\frac{\pi }{8},0\} $. To ensure the safety of the agent during operation, $\theta$ is selected from values within the range $\left[\frac{\pi}{4}, \frac{\pi}{2}\right]$, specifically $\left\{ {\frac{\pi }{4},\frac{{3\pi }}{{16}},\frac{\pi }{8},\frac{\pi }{{16}},0} \right\}$. The PEG trajectories of the agents are shown in Fig. \ref{morl-fql-guiji}. It can be observed that as $\varphi$ gradually decreases from $\frac{\pi}{2}$, the algorithm's preference for the agent's objective 2 (reaching target) diminishes, while its preference for objective 1 (evading pursuit) increases. Consequently, the success rate of the agent reaching the target drops, and it shifts to evading pursuit. However, since the speed of the pursuer is higher than that of the evader, the evader gets captured. As $\theta$ gradually decreases from $\frac{\pi}{4}$, the algorithm's preference for the agent's obstacle avoidance objective increases. This leads the agent to choose a path that is farther away from the obstacle, thus reducing the success rate of the pursuer in capturing the evader.

\begin{figure}[!ht]
	\centering
	\includegraphics[width=1.1\columnwidth]{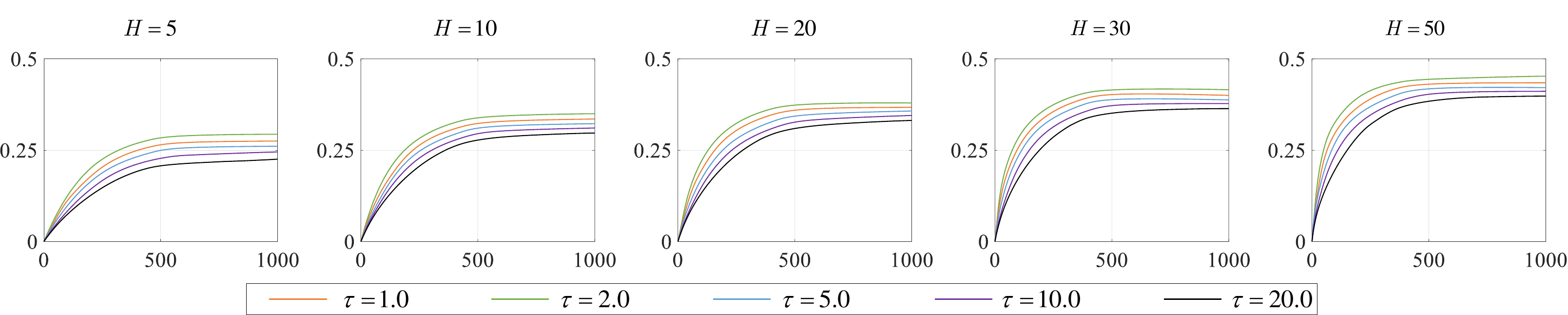}
	\caption{Global hypervolume with different $H$ and temperature $ \tau $.}
	\label{FQL-tau-H}
\end{figure}

To verify the impact of the temperature coefficient $\tau$ of the softmax function in \eqref{Pr-softmax} on the algorithm performance under different sampling numbers $H$, five different values of $H = 5, 10, 20, 30, 50$ are set, and the changing trends of the global strategy hypervolume under five different values of $\tau = 1.0, 2.0, 5.0, 10.0, 20.0$ are analyzed. To increase the exploration and diversity of strategy selection, a relatively high temperature coefficient should be chosen. However, as shown in Fig. \ref{FQL-tau-H}, it can be seen that when $\tau = 2.0$, the global hypervolume is the largest. Therefore, a higher temperature coefficient is not always better. Setting too high a temperature coefficient may prevent the model from quickly converging to the optimal solution, degrade the model's performance, and even lead to instability in the training process. In addition, under different $\tau$ values, as $H$ increases, the global hypervolume will increase. 

\begin{figure}[!ht]
	\centering
	\includegraphics[width=1.1\columnwidth]{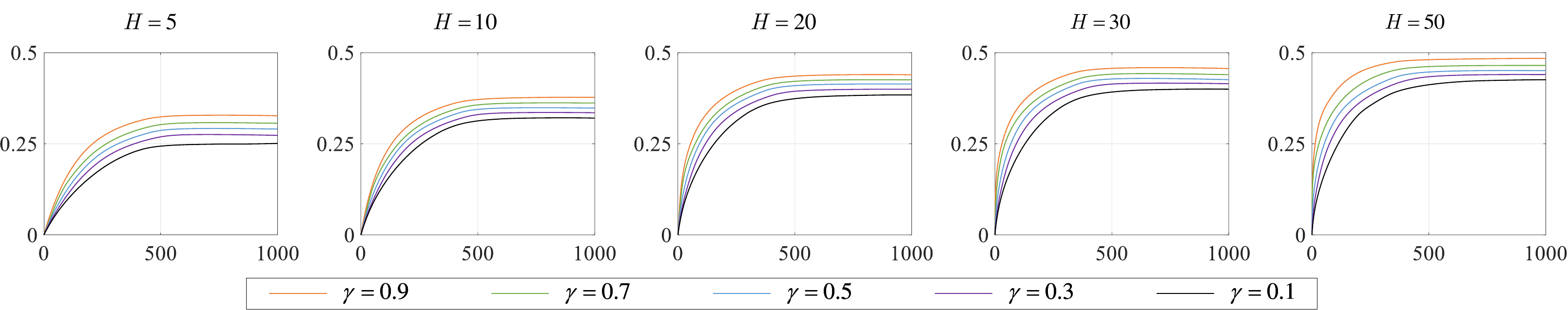}
	\caption{Global hypervolume with different $H$ and discount factors $\gamma$.}
	\label{FQL-guiji3}
\end{figure}

To verify the impact of the discount factor $\gamma$ on the algorithm performance under different sampling numbers $H$, five different values of $H = 5, 10, 20, 30, 50$ are set, and the changing trends of the global strategy hypervolume under five different values of $\gamma = 0.9, 0.7, 0.5, 0.3, 0.1$ are analyzed. As shown in Fig. \ref{FQL-guiji3}, it can be seen that as $\gamma$ increases, the global hypervolume also increases accordingly. This is because as $\gamma$ increases, the algorithm can obtain more future rewards. At the same time, as $H$ increases, the global hypervolume also increases. This is because a larger $H$ means that the algorithm can obtain more solutions on the Pareto front, enabling the agent to execute better strategies.

\begin{figure}[!ht]
	\centering
	\includegraphics[width=0.95\columnwidth]{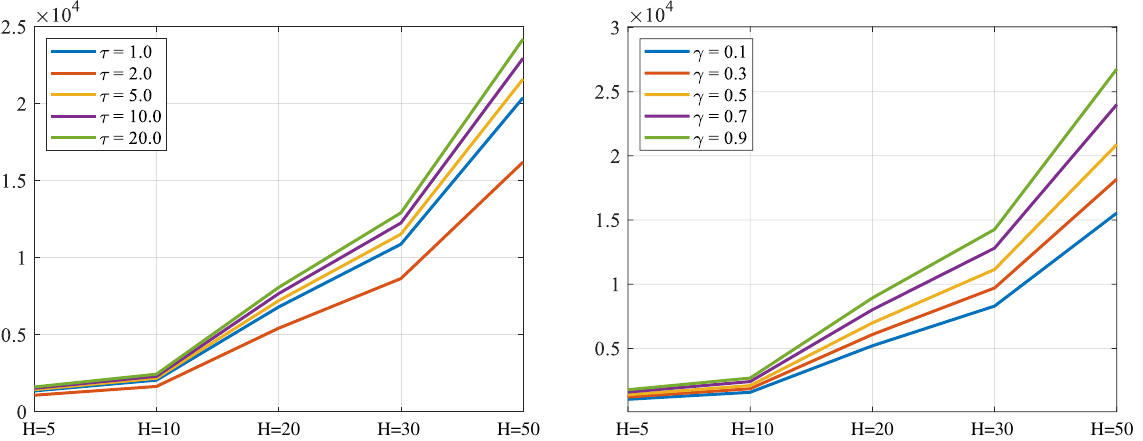}
	\caption{The average simulation time of the PEG under different conditions.}
	\label{H-tau-gamma}
\end{figure}

Table \ref{tab-time} presents the simulation time of PEGs under different sampling numbers $H$, different temperature coefficients $\tau$ and different discount factors $\gamma$. Fig. \ref{H-tau-gamma} illustrates the curves of the average simulation time of the computer as it changes with $H$ under different values of $\tau$ and $\gamma$. It can be observed that when the temperature coefficient $\tau$ takes the value of $2.0$, the average simulation time reaches its minimum. This phenomenon indicates that a larger value of the temperature coefficient does not invariably lead to better performance. As $\gamma$ increases, the simulation time gradually increases because the algorithm needs to calculate and store more cumulative rewards. As $H$ increases, the simulation time approximately increases multiplicatively.

\begin{table}[!ht]
        \label{tab-time}
        \caption{Simulation time of PEGs ($s$)}
        \begin{center}
            \tabcolsep 12pt
            \begin{tabular}{cccccc}\hline
		$H=5$ &$\tau = 1.0$ & $\tau = 2.0$ & $\tau = 5.0$ & $\tau = 10.0$ & $\tau = 20.0$\\ \hline
		$\gamma = 0.1$& 1002 & 884  & 1082 & 1090 & 1113\\
		$\gamma = 0.3$& 1122 & 944  & 1256 & 1309 & 1425\\
		$\gamma = 0.5$& 1368 & 1023 & 1407 & 1543 & 1613 \\
		$\gamma = 0.7$& 1595 & 1149 & 1628 & 1772 & 1846 \\
		$\gamma = 0.9$& 1702 & 1395 & 1821 & 1931 & 2061 \\ 
		$H=10$ &$\tau = 1.0$ & $\tau = 2.0$ & $\tau = 5.0$ & $\tau = 10.0$ & $\tau = 20.0$\\ \hline
		$\gamma = 0.1$& 1552 & 1374  & 1646 & 1671 & 1714\\
		$\gamma = 0.2$& 1684 & 1458  & 1901 & 1973 & 2157\\
		$\gamma = 0.5$& 2091 & 1581  & 2119 & 2347 & 2463 \\
		$\gamma = 0.7$& 2395 & 1753  & 2492 & 2692 & 2817 \\
		$\gamma = 0.9$& 2560 & 2103  & 2756 & 2940 & 3101 \\ 
		$H=20$ &$\tau = 1.0$ & $\tau = 2.0$ & $\tau = 5.0$ & $\tau = 10.0$ & $\tau = 20.0$\\ \hline
		$\gamma = 0.1$& 5038 & 4454 & 5443 & 5464 & 5598\\
		$\gamma = 0.3$& 5657 & 4738 & 6328 & 6580 & 7133\\
		$\gamma = 0.5$& 6879 & 5136 & 7059 & 7753 & 8079 \\
		$\gamma = 0.7$& 8021 & 5763 & 8165 & 8889 & 9238 \\
		$\gamma = 0.9$& 8557 & 7010 & 9152 & 9663 & 10330 \\ 
		$H=30$ &$\tau = 1.0$ & $\tau = 2.0$ & $\tau = 5.0$ & $\tau = 10.0$ & $\tau = 20.0$\\ \hline
		$\gamma = 0.1$& 8049 & 7114 & 8665 & 8763 & 5597\\
		$\gamma = 0.3$& 8988 & 7577 & 10067 & 10485 & 11427\\
		$\gamma = 0.5$& 10952 & 8187 & 11265 & 12380 & 12948 \\
		$\gamma = 0.7$& 12776 & 9220 & 13036 & 14219 & 14786 \\
		$\gamma = 0.9$& 13653 & 11199 & 14590 & 15489 & 16504 \\
		$H=50$ &$\tau = 1.0$ & $\tau = 2.0$ & $\tau = 5.0$ & $\tau = 10.0$ & $\tau = 20.0$\\ \hline
		$\gamma = 0.1$& 15044 & 13292 & 16232 & 16356 & 16699\\
		$\gamma = 0.3$& 16839 & 14184 & 18865 & 19638 & 21409\\
		$\gamma = 0.5$& 20521 & 15355 & 21123 & 23161 & 24225 \\
		$\gamma = 0.7$& 23969 & 17242 & 24421 & 26618 & 27736 \\
		$\gamma = 0.9$& 25551 & 20945 & 27358 & 28994 & 30994 \\ \hline
	\end{tabular}
        \end{center}  
\end{table}

%%%%%%%%%%%%%%%%%%%%%%%%%%%%%%%%%%%%
\section{Conclusion}\label{5}
For the PEG problem with multi-objective optimization considered in a continuous environment, this paper proposes a multi-objective FQL algorithm. Three optimization objectives, including evading pursuit, reaching target, and avoiding obstacle, are described through the reward functions. A multi-objective evaluation method based on three-dimensional hypervolume and an action selection strategy are designed. The computational complexity is reduced by approximately sampling the Pareto front. Different sampling angles are used to represent the preference degrees for different optimization objectives, and a global strategy update rule is designed. The performance of the proposed algorithm is verified through simulation results.

\end{document}